\documentclass[12pt]{article}

\usepackage{epsf}
\usepackage[dvips]{graphicx}

\def\GeV{\,\mbox{GeV}}

\def\lsim{\mathrel{\rlap{\lower4pt\hbox{\hskip1pt$\sim$}}
    \raise1pt\hbox{$<$}}}         
\def\gsim{\mathrel{\rlap{\lower4pt\hbox{\hskip1pt$\sim$}}
    \raise1pt\hbox{$>$}}}         
\def\BA{\begin{eqnarray}}
\def\BE{\begin{equation}}
\def\BF{\begin{figure}[htb]}
\def\BT{\begin{table}[htb]}
\def\EA{\end{eqnarray}}
\def\EE{\end{equation}}
\def\EF{\end{figure}}
\def\ET{\end{table}}

\def\la{\langle}
\def\ra{\rangle}

\begin{document}

\begin{center}
{\bfseries PARTON RESCATTERINGS

IN LARGE-$x$ NUCLEAR SUPPRESSION AT RHIC}

\vskip 5mm

J.~Nemchik$^{1,2}$ and
        M.~\v Sumbera$^{3 \dag}$

\vskip 5mm

{\small
(1)
{\it Institute of Experimental Physics SAS, Watsonova 47,
04001 Ko\v sice, Slovakia}
\\
(2) {\it Czech Technical University in Prague, FNSPE, B\v rehov\'
a 7, 11519 Praque, Czech Republic}
\\
(3) {\it Nuclear Physics Institute AS CR,
25068 \v Re\v z/Prague, Czech Republic} \\
$\dag$ {\it speaker, E-mail: sumbera@ujf.cas.cz}}
\end{center}

\vskip 5mm

\begin{center}
\begin{minipage}{150mm}
\centerline{\bf Abstract}

We demonstrate that strong suppression of the relative production
rate $(d+Au)/(p+p)$ of inclusive high-$p_T$ hadrons at forward
rapidities observed at RHIC is due to parton multiple
rescatterings in nuclear matter. The light-cone dipole
approach-based calculations are in a good agreement with BRAHMS
and STAR data. They also indicate a significant nuclear
suppression at midrapidities with a weak onset of the coherence
effects. This prediction is supported by the preliminary $d+Au$
data from the PHENIX Collaboration. Moreover, since similar
suppression pattern is also expected to show up at lower energies
where effects of parton saturation are not expected, we are able
to exclude from the interpretation of observed phenomena models
based on the Color Glass Condensate.

\end{minipage}
\end{center}

\vskip 10mm

%
%
\section{Introduction}
\label{intro}
%
%

Spectra of high-$p_T$ hadrons produced in nuclear collisions at
large forward rapidities are promising tool to study partonic
degrees of freedom in nuclei. Strong nuclear suppression of the
spectra observed by the BRAHMS \cite{brahms,brahms-07} and STAR
\cite{star} Collaborations in deuteron-gold collisions at the
Relativistic Heavy Ion Collider (RHIC) was a tempting invitation
for the parton saturation \cite{glr,al} or the Color Glass
Condensate (CGC) \cite{mv} motivated phenomenology \cite{kkt} as
its most natural explanation. \vspace*{-0.075cm}

According to these models the parton coherence phenomena may
reveal itself already at RHIC energies showing up first in the
wave function of heavy nuclei. Kinematically most favorable region
to access the strongest coherence effects is the fragmentation
region of the light nucleus $1$ colliding with the heavy one $2$.
At large $x_1$ (large Feynman $x_F$ at forward rapidities) one can
reach the smallest values of the light-front momentum fraction
variable $x_2=x_1-x_F$ ($2\times 10^{-4}\lsim x_2\lsim 10^{-3}$ in
the RHIC kinematic range). \vspace*{-0.075cm}

Quite unexpectedly, the same nuclear effects occur not only at
forward rapidities \cite{brahms,brahms-07,star} but also in the
large $p_T$ region at midrapidity \cite{phenix} where effects of
coherence are not important. The covered interval of $x_2\gsim
0.01$ goes too far beyond the region where the CGC is valid.
\vspace*{-0.075cm}

In \cite{knpsj-05,npps-08} it was shown  that for any large-$x_1$
reaction considerable nuclear suppression comes from the energy
conservation at the level of projectile partons undergoing
multiple rescatterings in nuclear medium. It was also demonstrated
\cite{knpsj-05} that large-$x_1$ suppression is a leading twist
effect, violating QCD factorization, a basic ingredient of the
CGC-based models. \vspace*{-0.075cm}

Analysis of nuclear suppression based on the multiple parton
rescatterings leads also to a new type of scaling: the same
nuclear effect are expected at different energies and rapidities
corresponding to the same value of $x_1$ ($x_F$ at forward
rapidities) \cite{knpsj-05,npps-08}. The most straight forward
prediction of the $x_1$-scaling is that similar nuclear effects
must also show up at lower c.m. energy $\sqrt s$. Here the onset
of coherence effects is much weaker and so there is much less room
for explanation of strong nuclear suppression in terms of the CGC.

Another consequence of this scaling is that in the RHIC energy
range similar nuclear effects must also show up at midrapidities
provided that the corresponding values of $p_T$ of produced
hadrons reach the same value of $x_1$ as at forward rapidities.
This prediction is confirmed by the preliminary data on neutral
pion production in $d+Au$ collisions measured recently by the
PHENIX experiment \cite{phenix} showing an evidence for the
nuclear suppression at rather large $p_T > 8\,$GeV. This and new
2008 $d+Au$ high-statistic data may provide another test of our
approach.

%
%
\section{High-$p_T$ hadron production: Sudakov suppression,
production cross section}
\label{scatt}
%
%

Let us recall that in the limit $x_1\to 1$ ($x_F\to 1$ at forward
rapidities) gluon radiation in any pQCD-driven hard scattering is
forbidden by the energy conservation. For uncorrelated Poisson
distribution of radiated gluons, the Sudakov suppression factor,
i.e. the probability to have a rapidity gap $\Delta y =
-\ln(1-x_1)$ between leading parton and rest of the system, has a
very simple form: $S(x_1) = 1-x_1$ \cite{knpsj-05}.

Suppression at $x_1\to 1$ can thus be formulated as a survival
probability of the large rapidity gap (LRG) process in multiple
interactions of projectile valence quarks with the nucleus. Every
additional inelastic interaction of the quarks contributes an
extra suppression factor $S(x_1)$. The probability of an n-fold
inelastic collision is related to the Glauber model coefficients
via the Abramovsky-Gribov-Kancheli (AGK)  cutting rules
\cite{agk}. Correspondingly, the survival probability at impact
parameter $\vec b$ reads
%
%
 \BE
W^{hA}_{LRG}(b) =
\exp[-\sigma_{in}^{hN}\,T_A(b)]\,
\sum\limits_{n=1}^A\frac{1}{n!}\,
\left[\sigma_{in}^{hN}\,T_A(b)\,
\right]^n\,S(x_1)^{n-1}\ ,
 \label{70}
 \EE
%
%
where $T_A(b)$ is the nuclear thickness function.

At large $p_T$, the cross section of hadron production in
$d+A\,(p+p)$ collisions is given by a convolution of the
distribution function for the projectile valence quark with the
quark scattering cross section and the fragmentation function
%
%
 \BE
\frac{d^2\sigma}{d^2p_T\,d\eta} =
\sum\limits_q \int\limits_{z_{min}}^1 dz\,
f_{q/d(p)}(x_1,q_T^2)\,
\left.\frac{d^2\sigma[qA(p)]}{d^2q_T\,d\eta}
\right|_{\vec q_T=\vec p_T/z}\,
\frac{D_{h/q}(z)}{z^2},
\label{80}
 \EE
%
%
where $x_1=\frac{q_T}{\sqrt{s}}\,e^\eta$.
For the quark distribution functions in the nucleon we use the
lowest order parametrization from \cite{grv}. Fragmentation
functions were taken from \cite{fs-07}.

As first shown in \cite{knpsj-05,npps-08} the effective projectile
quark distribution correlates with the target. So interaction with
the nuclear target does not obey the factorization. Main source of
suppression at large $p_T$ comes from multiple soft rescatterings
of the quark in nuclear matter. Summed over multiple interactions,
the quark distribution in the nucleus reads
%
%
 \BE
\hspace*{-0.40cm}
f^{(A)}_{q/N}(x_1,q_T^2) = C\,f_{q/N}(x_1,q_T^2)\,
\frac{\int d^2b\,
\left[e^{-x_1\sigma_{eff}T_A(b)}-
e^{-\sigma_{eff}T_A(b)}\right]}
{(1-x_1)\int d^2b\,\left[1-
e^{-\sigma_{eff}T_A(b)}\right]}\, ,
\label{100}
 \EE
%
%
where effective cross section $\sigma_{eff} =
\sigma_{eff}(p_T,s)=\frac{\left\la \sigma^2_{\bar
qq}(r_T)\right\ra} {\left\la \sigma_{\bar qq}(r_T)\right\ra}$ has
been evaluated in \cite{knpsj-05}. The normalization factor $C$ in
Eq.~(\ref{100}) is fixed by the Gottfried sum rule.

The cross section of quark scattering on the target
$d\sigma[qA(p)]/d^2q_Td\eta$ in Eq.~(\ref{80}) is calculated in
the light-cone dipole approach \cite{zkl,jkt-01}. We separate
contributions with different initial transverse momenta and sum
over different mechanisms of high-$p_T$ hadron production. Details
can be found in \cite{knpsj-05}.

Let us note that in the RHIC energy range and at midrapidity
correct description of hadrons with small and moderate $p_T$  can
be achieved only if the above calculations incorporate production
and fragmentation of gluons\cite{knst-01}. Consequently, the cross
section for hadron production, Eq.~(\ref{80}), should be extended
also for gluons with corresponding distribution function, parton
scattering cross section and the fragmentation function. Including
multiple parton interactions, the gluon distribution in the
nucleus is given by the same formula as for quarks (see
Eq.~(\ref{100})), except $\sigma_{eff}$, which should be
multiplied by the Casimir factor $9/4$.

%
%
\section{Comparison with data}
\label{data}
%
%

\begin{figure}[tbh]
\epsfysize=50mm
\resizebox{18.6pc}{!}{
    \includegraphics[height=0.2\textheight]{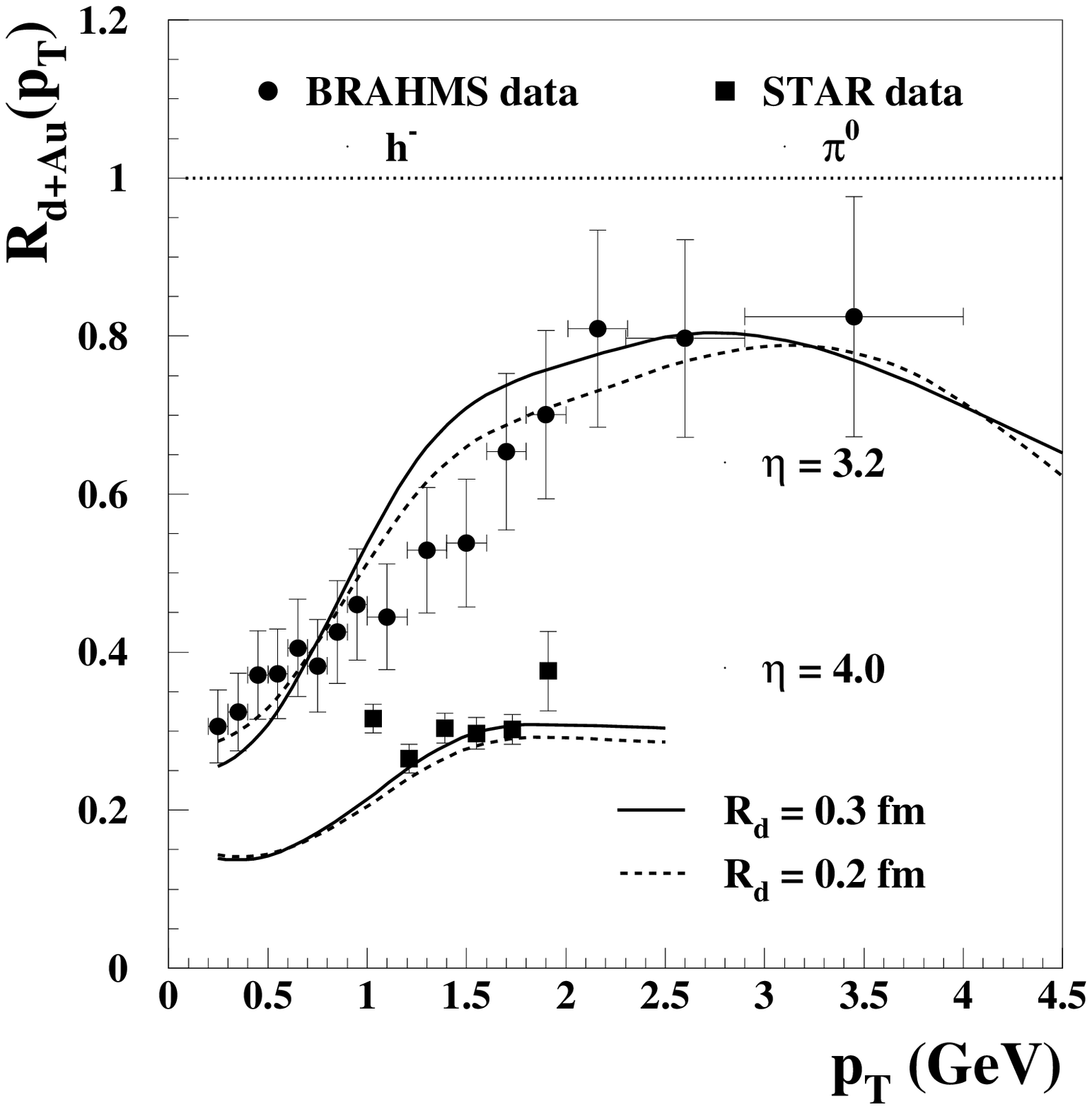}}
\hspace*{0.8cm}
\resizebox{18.3pc}{!}{
     \includegraphics[height=0.4\textheight]{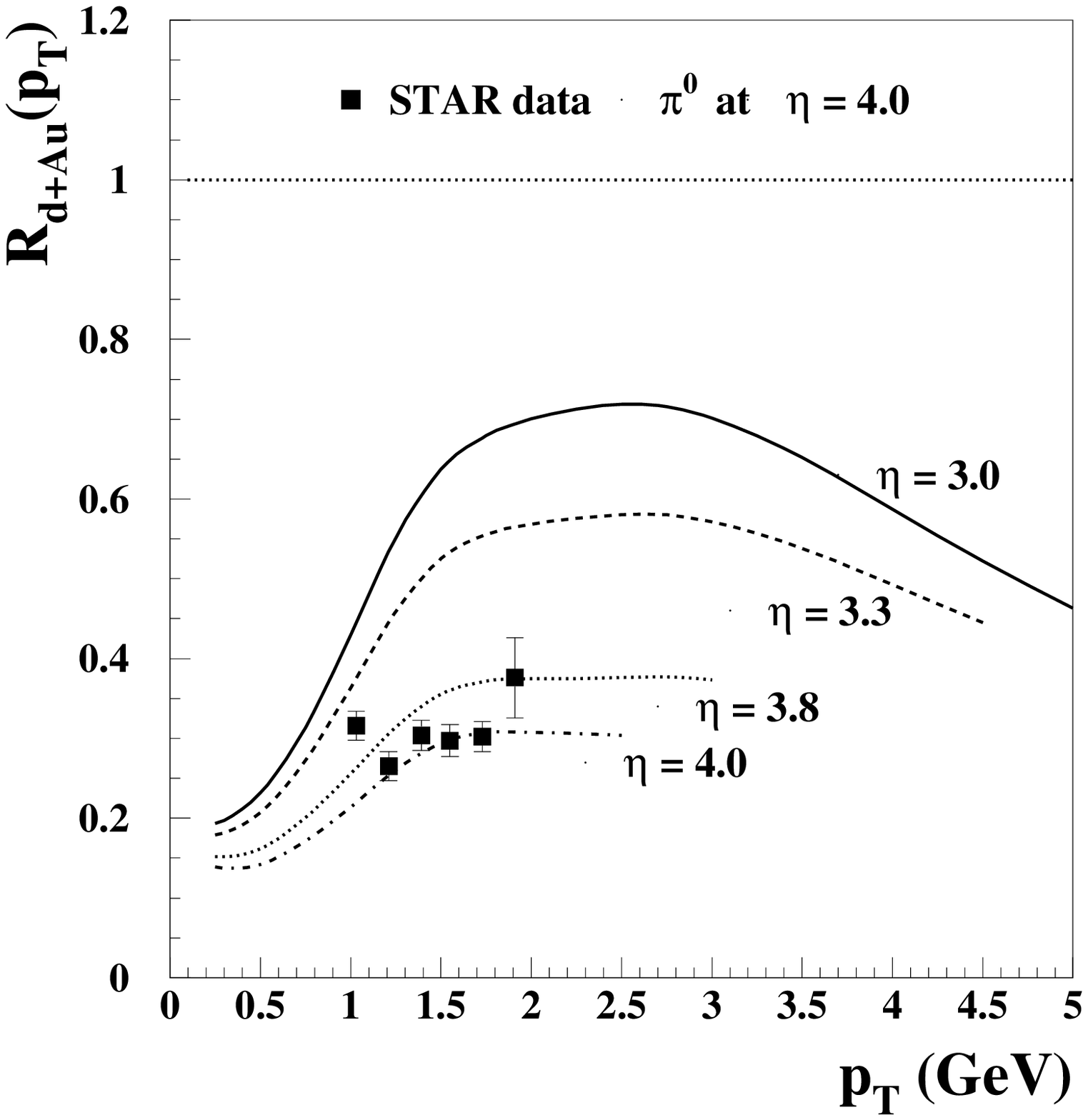}}
\vspace*{-0.7cm}
\caption
 {\small
(Left) Ratio of negative hadron and neutral pion production rates
in $d+Au$ and $p+p$ collisions as function of $p_T$ at $\eta =
3.2$ and $\eta = 4.0$. Data are from the BRAHMS \cite{brahms} and
STAR Collaborations \cite{star}, respectively. (Right) Model
predictions for the ratio $R_{d+Au}(p_T)$ for production of
$\pi^0$ mesons at $\sqrt{s}=200\GeV$ and different values of
$\eta$ changing from $3$ to $4$. } \label{star}
 \end{figure}

In 2004 the BRAHMS Collaboration \cite{brahms} observed a
significant nuclear suppression of negative hadrons produced at
$\eta=3.2$. Much stronger onset of nuclear effects was observed
later on by the STAR Collaboration \cite{star} for $\pi^0$
production at pseudorapidity $\eta = 4.0$. Both measurements are
plotted in the left panel of Fig.~\ref{star}. A huge difference in
nuclear suppresion factor at different $\eta$ is due to the energy
conservation and reflects much smaller survival probability of the
LRG in multiple parton interactions at larger $x_1$
\cite{knpsj-05,npps-08}.

To demonstrate different onsets of nuclear effects with increasing
pseudorapidity we present in the right panel of Fig.~\ref{star}
our calculations for the nuclear suppression factor at different
fixed values of $\eta$. Changing the value of $\eta$ from $3.0$ to
$4.0$ leads to a rise of $R_{d+Au}(p_T)$ by a factor of 2
\cite{npps-08}.

\begin{figure}[th]
\resizebox{17.3pc}{!}{
     \includegraphics[height=0.4\textheight]{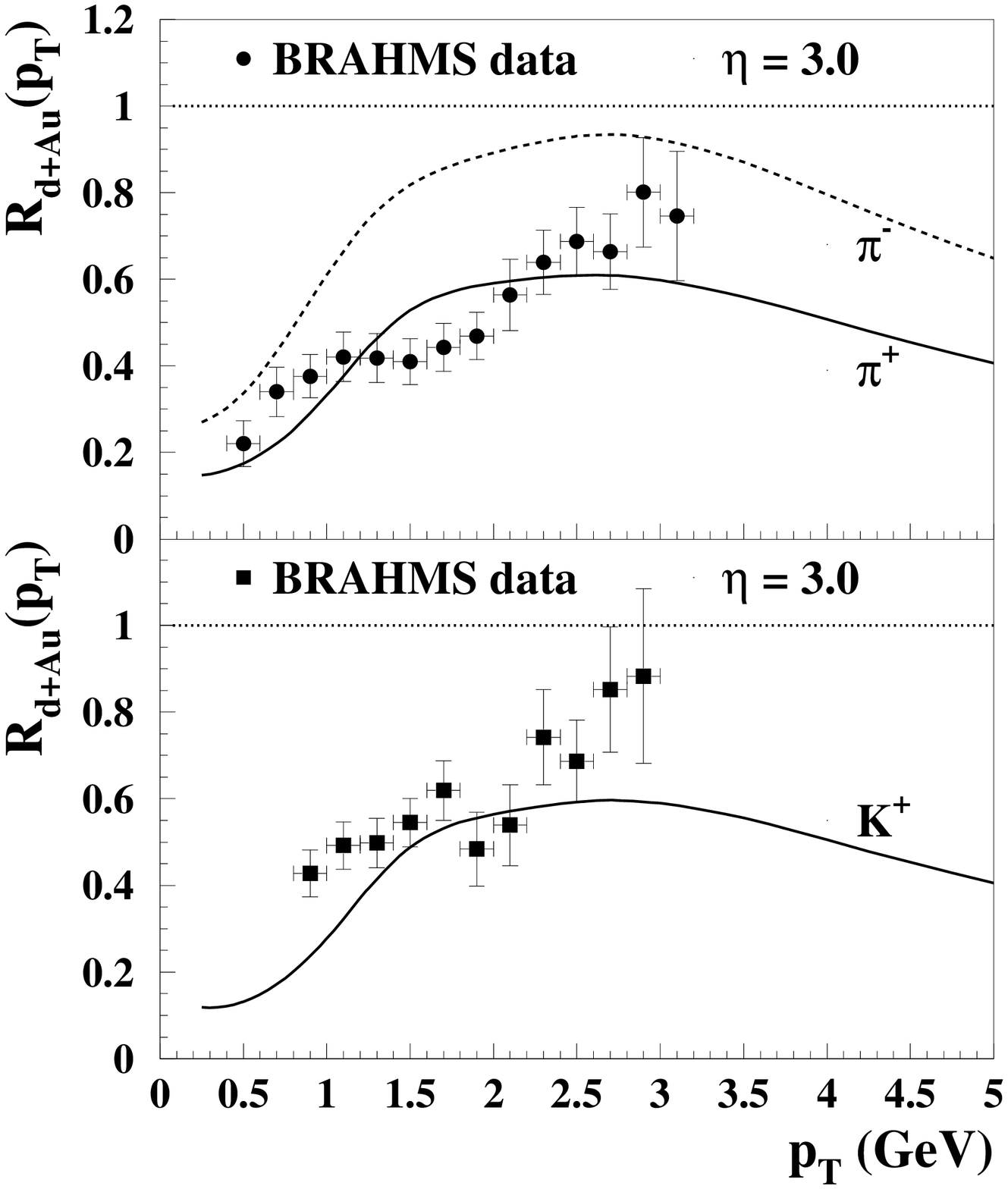}}
\hspace*{0.70cm}
\resizebox{19.8pc}{!}{
     \includegraphics[height=0.4\textheight]{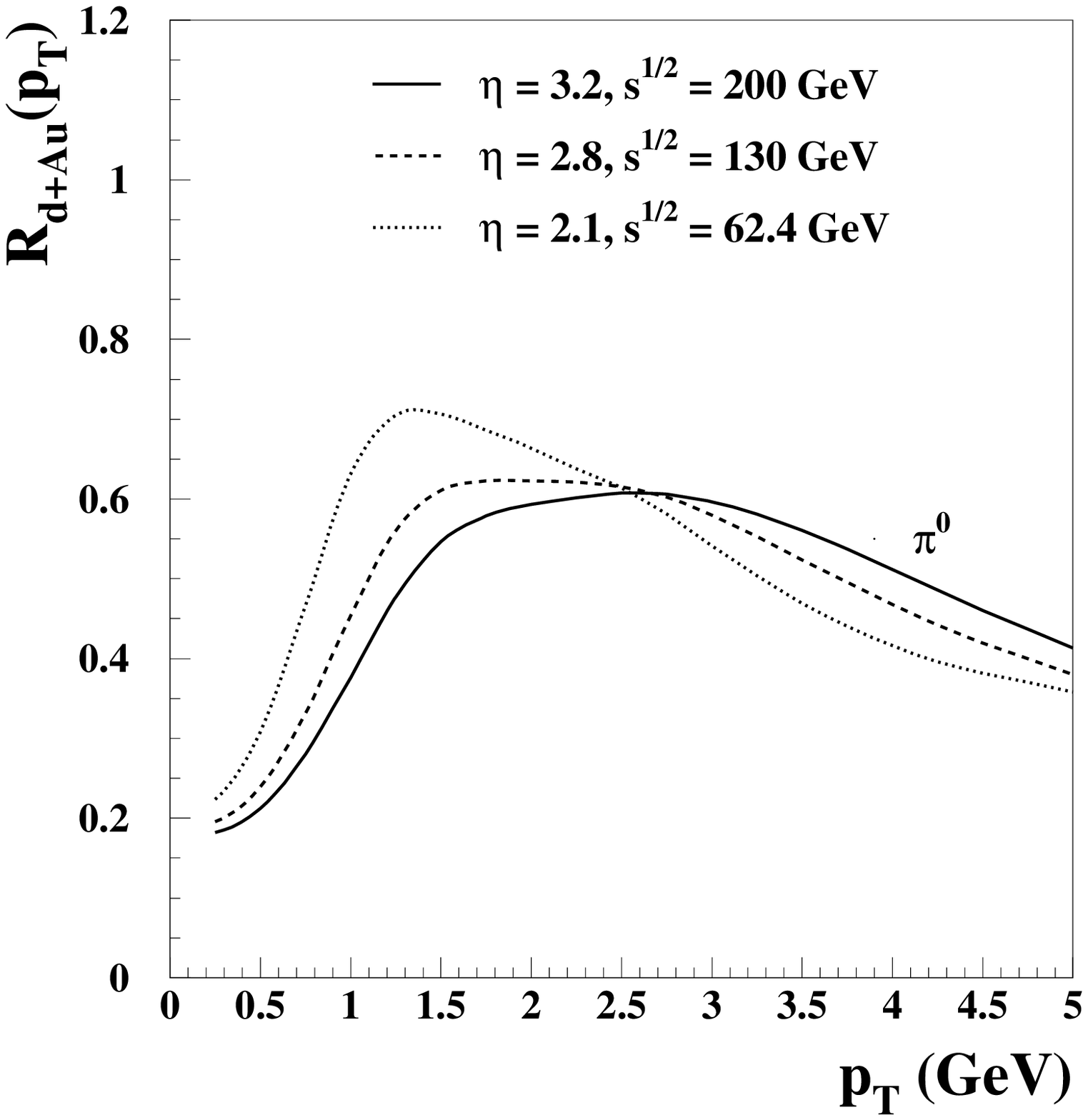}}
\vspace*{-0.2cm}
\caption
 {\small
(Left) Ratio, $R_{d+Au}(p_T)$ for identified particles produced in
$d+Au$ and $p+p$ collisions at $\eta = 3$. The data are from the
BRAHMS Collaboration \cite{brahms-07}. (Right) Predictions for the
ratio $R_{d+Au}(p_T)$ for $\pi^0$ at different $\eta$ and $\sqrt
s$ having the same $\exp(\eta)/\sqrt{s}$.}
 \label{scaling}
\end{figure}

The BRAHMS Collaboration has recently reported a new measurements
\cite{brahms-07} on production of positively charged pions and
kaons at $\eta = 3.0$ in $d+Au$ collisions confirming suppression
pattern they found in 2004 for the negative
particles\cite{brahms}. Their recent data are plotted on the left
panel of Fig.~\ref{scaling} together with our model predictions.

The calculations of $R_{d+Au}(p_T)$ of neutral pions at $\sqrt{s}
= 200, 130$ and $62.4\,$GeV shown on the right panel of
Fig.~\ref{scaling} reveal approximate $x_1$($x_F$)-scaling at RHIC
energy range, i.e. the same nuclear effects at values of $\eta$
and $\sqrt s$ corresponding to the same value of $x_1$.

\begin{figure}[th]
\resizebox{18.5pc}{!}{
     \includegraphics[height=0.4\textheight]{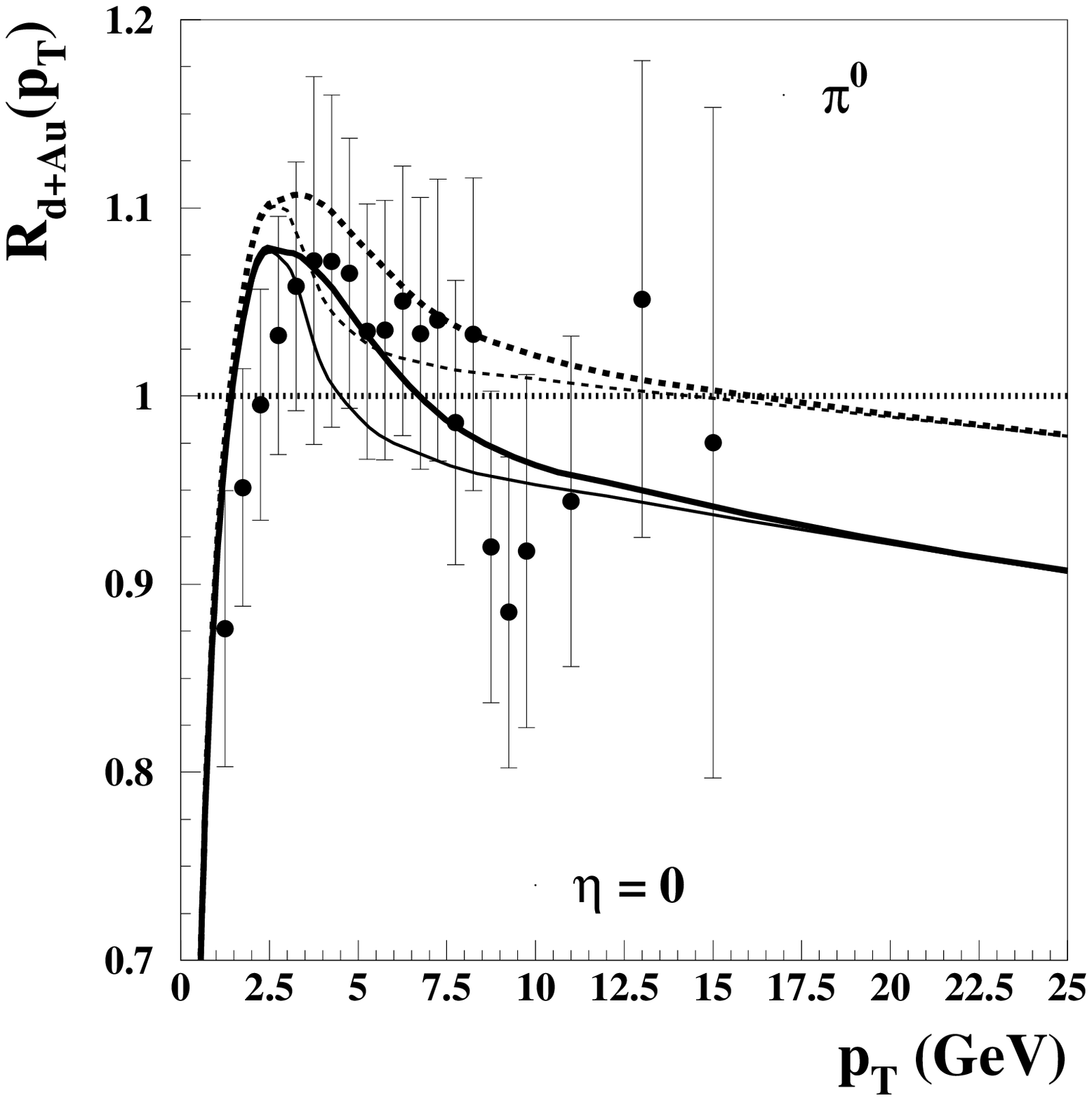}}
\hspace*{0.80cm}
\resizebox{18.5pc}{!}{
     \includegraphics[height=0.4\textheight]{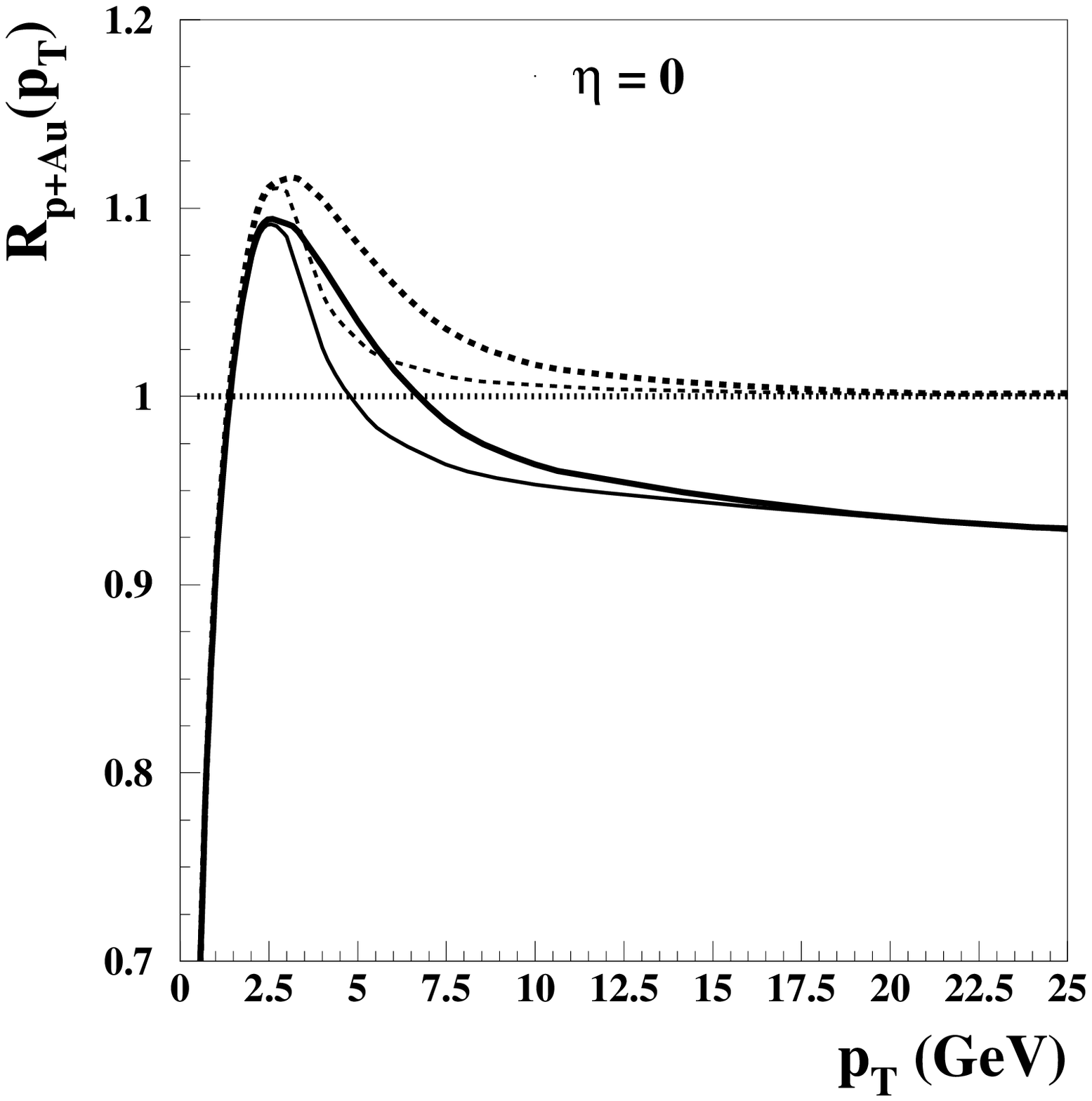}}
\vspace*{-0.8cm}
\caption
 {\small
(Left)
Ratio $R_{d+Au}(p_T)$ as a function of
$p_T$ for production of $\pi^0$ mesons
at $\sqrt{s}=200\GeV$ and $\eta = 0$ vs. data from the
PHENIX Collaboration \cite{phenix}. Thin solid and dashed lines
represent the predictions calculated in the limit
of long coherence length. Thick solid and dashed lines
include corrections for the finite coherence length.
(Right)
The same as Fig. in the left panel but for the ratio $R_{p+Au}(p_T)$.
}
 \label{phenix}
\end{figure}

Generalization of the $x_1$-scaling from the forward region to
midrapidity is studied on Fig.~\ref{phenix}. The only difference
to the previous analysis is that the same value of $x_1$ at
midrapidity as that in the forward region requires substantially
higher hadron transverse momenta. On the left panel of
Fig.~\ref{phenix} our predictions for the nuclear suppression
factor of $\pi^0 $ produced in $d+Au$ collisions at midrapidities
are confronted with the recent data of the PHENIX
Collaboration\cite{phenix}.
%
%

Here the thin dashed line corresponds to the case when multiple
parton rescatterings are not taken into account. The calculations
with inclusion of multiple parton rescatterings are presented by
the thin solid line. At moderate $p_T\in (3,7)\,$GeV the model
underestimates the data. However, quite a strong onset of nuclear
suppression at large $p_T$ is not in a disagreement with
corresponding experimental points. At $p_T = 25\,$GeV we expect
$R_{d+Au}(p_T) \sim 0.9$.

Due to the transition between the regimes with (small $p_T$) and
without (large $p_T$) onset of coherence effects in the RHIC
energy range calculations at $\eta = 0$ are very complicated. One
can deal with this situation relying on the light-cone Green
function formalism \cite{knst-02,krt2,n-08} but the integrations
involved become too complicated. To simplify the situation we have
used instead corrections for finite coherence length. Following
the procedure described in\cite{knst-01} we have used linear
interpolation performed by the means of so-called nuclear
longitudinal form factor. Such a situation is shown by the thick
solid and dashed lines on Fig.~\ref{phenix} corresponding to the
case with and without inclusion of the multiple parton
rescatterings, respectively. One can see that this correction
brings the model predictions to a better agreement with the data
at moderate $p_T$.

On the right panel of Fig.~\ref{phenix} we also present model
predictions for the ratio $R_{p+Au}$ as a function of $p_T$.
Compared to $d+Au$ system study of nuclear effects in $p+Au$
minimizes the isospin effects. At $p_T = 25\,$GeV we predict
$R_{p+Au}\sim 0.93$.


%
%
\section{Summary and conclusions}\label{conclusions}
%
%

In the present paper we have analyzed consequences of the $x_1$
($x_F$)-scaling of the nuclear suppression factor $R_{p(d)+Au}$ of
high-$p_T$ hadrons at RHIC.

The new results are:
\begin{itemize}
\vspace*{-0.1cm}

 \item According to the $x_1$-scaling,
considerable nuclear suppression at large $x_1$
is expected for different kinematic regions :\\
- {\bf production of high-$p_T$ hadrons at forward rapidities.} \\
- {\bf production of high-$p_T$ hadrons
at smaller rapidities and smaller energies}. \\
- {\bf productions of hadrons with very large $p_T$ at midrapidity.}

\item Using simple formula (\ref{100}) adopted from
\cite{knpsj-05} and based on the Glauber multiple interaction
theory and the AGK cutting rules, we have calculated high-$p_T$
hadron production at midrapidity and found quite a strong nuclear
suppression. This observation does not contradict to the recent
measurements of the PHENIX Collaboration \cite{phenix}.

\item In order to avoid the isospin effects, we have also studied
large-$p_T$ neutral pion production in $p+Au$ collisions. With the
same input, we predict (see the right panel of Fig.~\ref{phenix})
for the first time quite a strong nuclear suppression, $R_{p+Au} =
0.93$ at $p_T = 25\,$GeV.

\item In the RHIC kinematic region, investigation of large-$x$
 hadron production in $p(d)+Au$ collisions at midrapidities
 represents the baseline for verification of different
 phenomenological models.
important. At high-$p_T$ the data cover region of $x_2\sim
0.05-0.1$ where effects of coherence are negligible allowing to
exclude the CGC-based models from interpretation of observed
nuclear suppression.

\end{itemize}

{\bf Acknowledgments} This work was supported in part by the Grant
Agency of the Czech Republic, Grant 202/07/0079, Slovak Funding
Agency, Grant 2/7058/27; and by Grants VZ MSM 6840770039 and LC
07048 (Ministry of Education of the Czech Republic).
\vspace*{-0.2cm}



\begin{thebibliography}{99}

\bibitem{brahms}
BRAHMS Collaboration, I.~Arsene {\em et al.},
Phys. Rev. Lett. {\bf 93}, 242303 (2004).

\bibitem{brahms-07}
BRAHMS Collaboration, Hongyan Yang {\em et al.},
J. Phys. {\bf G34}, S619 (2007).

\bibitem{star}
STAR Collaboration, J.~Adams {\em et al.},
Phys. Rev. Lett. {\bf 97}, 152302 (2006).

\bibitem{glr}
L.V.~Gribov, E.M.~Levin, and M.G.~Ryskin,
Nucl. Phys. {\bf B188}, 555 (1981);
Phys. Rep. {\bf 100}, 1 (1983).

\bibitem{al}
A.H.~Mueller,
Eur. Phys. J. {\bf A1}, 19 (1998).

\bibitem{mv}
L.~McLerran, and R.~Venugopalan, Phys. Rev. {\bf D49}, 2233
(1994); ibid, 3352.

\bibitem{kkt}
D.~Kharzeev, Y.V.~Kovchegov, and K.~Tuchin,
Phys. Lett. {\bf B599}, 23 (2004).

\bibitem{phenix}
PHENIX Collaboration,
S.S.~Adler et al.,
Phys. Rev. Lett. {\bf 98}, 172302 (2007).

\bibitem{knpsj-05}
B.Z.~Kopeliovich, J.~Nemchik, I.K.~Potashnikova, I.~Schmidt,
and
M.B.~Johnson,
Phys. Rev. {\bf C72}, 054606 (2005).

\bibitem{npps-08}
J.~Nemchik, V.~Petr\'a\v cek, I.K.~Potashnikova, and M.~\v
Sumbera, Phys. Rev. {\bf C78}, 025213 (2008).

\bibitem{agk}
A.V.~Abramovsky, V.N.~Gribov, and O.V.~Kancheli,
Yad. Fiz. {\bf 18}, 595 (1973).

\bibitem{grv}
M.~Gluck, E.~Reya, and A.~Vogt,
Z. Phys. {\bf C67}, 433 (1995).

\bibitem{fs-07}
D.~de~Florian, R.~Sassot, and M.~Stratmann,
Phys. Rev. {\bf D75}, 114010 (2007);
Phys. Rev. {\bf D76}, 074033 (2007).

\bibitem{zkl}
A.B.~Zamolodchikov, B.Z.~Kopeliovich, and L.I.~Lapidus,
{\sl Pis'ma Zh. Eksp. Teor. Fiz.} {\bf 33}, 612 (1981);
{\sl Sov. Phys. JETP Lett.} {\bf 33}, 595 (1981).

\bibitem{jkt-01}
M.B.~Johnson, B.Z.~Kopeliovich, and A.V.~Tarasov,
Phys. Rev. {\bf C63}, 035203 (2001).

\bibitem{knst-01}
B.Z.~Kopeliovich, J.~Nemchik, A.~Sch\"afer, and A.V.~Tarasov,
Phys. Rev. Lett. {\bf 88}, 232303 (2002).

\bibitem{knst-02}
B.Z.~Kopeliovich, J.~Nemchik, A.~Sch\"afer, and A.V.~Tarasov,
Phys. Rev. {\bf C88}, 035201 (2002).

\bibitem{krt2}
B.Z.~Kopeliovich, J.~Raufeisen, and A.V.~Tarasov,
Phys. Rev. {\bf C62}, 035204 (2000).

\bibitem{n-08}
B.Z.~Kopeliovich, J.~Nemchik, I.K.~Potashnikova, and
I.~Schmidt,
J. Phys. {\bf G35}, 115010 (2008).

\end{thebibliography}
\end{document}